# Hydrogenating VO$_2$ with protons in acid solution

Yuliang Chen[1+], Zhaowu Wang[2,3+], Shi Chen[1], Hui Ren[1], Liangxin Wang[1], Guobin Zhang[1], Yalin Lu[1], Jun Jiang[2]*, Chongwen Zou[1]*, Yi Luo[2]

[1]National Synchrotron Radiation Laboratory, University of Science and Technology of China, Hefei, 230026, China.

[2] Hefei National Laboratory for Physical Sciences at the Microscale, iChEM (Collaborative Innovation Center of Chemistry for Energy Materials), Hefei Science Center (CAS), and School of Chemistry and Materials Science, University of Science and Technology of China, Hefei, 230026, China.

[3] School of Physics and Engineering，Henan University of Science and Technology，Luoyang, Henan 471023, P. R. China

Corresponding Authors: jiangj1@ustc.edu.cn ; czou@ustc.edu.cn
+These two authors contributed equally to this paper.

**Hydrogenation is an effective way to tune material property[1-5]. Traditional techniques for doping hydrogen atoms into solid materials are very costly due to the need for noble metal catalysis and high-temperature/pressure annealing treatment or even high energy proton implantation in vacuum condition[5-8]. Acid solution contains plenty of freely-wandering protons, but it is difficult to act as a proton source for doping, since the protons always cause corrosions by destroying solid lattices before residing into them. Here we achieve a facile way to hydrogenate monoclinic vanadium dioxide ($VO_2$) with protons in acid solution by attaching suitable metal to it. Considering the Schottky contact at the metal/$VO_2$ interface, electrons flow from metal to $VO_2$ due to workfunction difference and simultaneously attract free protons in acid solution to penetrate, forming the hydrogens dopants inside $VO_2$ lattice. This metal-acid treatment constitutes an electron-proton co-doping strategy, which not only protects the $VO_2$ lattice from corrosion, but also causes pronounced insulator-to-metal transitions. In addition, the metal-acid induced hydrogen doping behavior shows a ripple effect, and it can spread contagiously up to wafer-size area (>2 inch) even triggered by a tiny metal particle attachment (~1.0mm). This will stimulate a new way of simple and cost-effective atomic doping technique for some other oxide materials.**

As a typical transition oxide, $VO_2$ shows a pronounced MIT behavior at the critical temperature near 68°C, accompanying by a sharp resistance change up to five orders of magnitude and dramatic infrared switching effect within sub-*ps* time scale[9-13]. Due to its peculiar characteristics across the phase transition, $VO_2$ material is promising for important applications including memory material[14,15], smart window[16,17] and ultra-fast optical switching device[18]. While the relatively high critical phase transition temperature greatly hinders its real applications, thus many efforts have been made to modulate the phase transition behavior[2,3,19-24]. Hydrogen doping is widely utilized for material property optimization, for instance, H-incorporation in solid crystal can bestow visible-light driving photocatalysis ability to $TiO_2$[5], and modulate metal–insulator transitions (MIT) in strongly correlated oxides such as $SmNiO_3$ and other correlated oxides[1,3,6]. Recent experiments[2,8] also observed that H-incorporations in M-$VO_2$ resulted in a very stable metallic phase at room temperature, giving excellent thermoelectric performance[25]. Interestingly, further injecting H into the lightly doped M-$VO_2$ created another insulating state at the heavily H-doping situation[3], making it a promising technique of controlling MIT in a reversible and consecutive way. However, for all of these hydrogenation treatments, the H-intercalations into crystal lattice are very costly because of the needs of noble metal (Au, Pt, Pd) catalysts or high-pressure/temperature annealing process. That is, the creation of single-atom based hydrogens and their injection into solid cost a large amount of energies.

In this work, we propose a facile route to hydrogenate M-$VO_2$ with protons in acid solution. It is known that some metallic oxides including $VO_2$ are easily dissolved in acid. Protons ($H^+$) with positive charges attack oxygen atoms in oxide, and soon break the crystal lattice by dragging oxygen into solution, through a traditional reaction of $VO_2 + 4H^+ \rightarrow V^{4+} + 2H_2O$. As shown in Fig. 1a, a 30 nm M-$VO_2$/$Al_2O_3$ (0001) epitaxial film grown by molecular beam epitaxy method[26] (Supplementary Information Fig. S1), was held by a plastic tweezers and put into a 2% $H_2SO_4$ acid solution. As expected, the yellowy $VO_2$ epitaxial film completely disappeared after 3 hours. In contrast, when we used a steel tweezers as shown in Fig. 1b, the same $VO_2$ film suddenly obtained excellent anti-corrosion ability, as it was nearly intact by 3 hours in the acid solution. Obviously, the magic trick is ascribed to the metal attachment of the steel tweezers. Scanning electron microscope (SEM) images in Fig. 1c show that the thickness of $VO_2$ film remained unchanged and the surface maintained almost the same grain-like morphologies even after 20 hours in acid solution. More convincingly, the trace element analysis in Fig. 1d found that the $V^{4+}$ cations concentration in solution increased from 0.11μg/ml to 1.82 μg/ml after immersing a $VO_2$ film without metal-attachement in acid from 30 minutes to 20 hours, while that of metal-acid treated sample kept very low $V^{4+}$ concentration at 0.03~0.06 μg/ml in 20 hours. These results demonstrated excellent anti-corrosion ability of $VO_2$ lattice in acid if metal was attached.

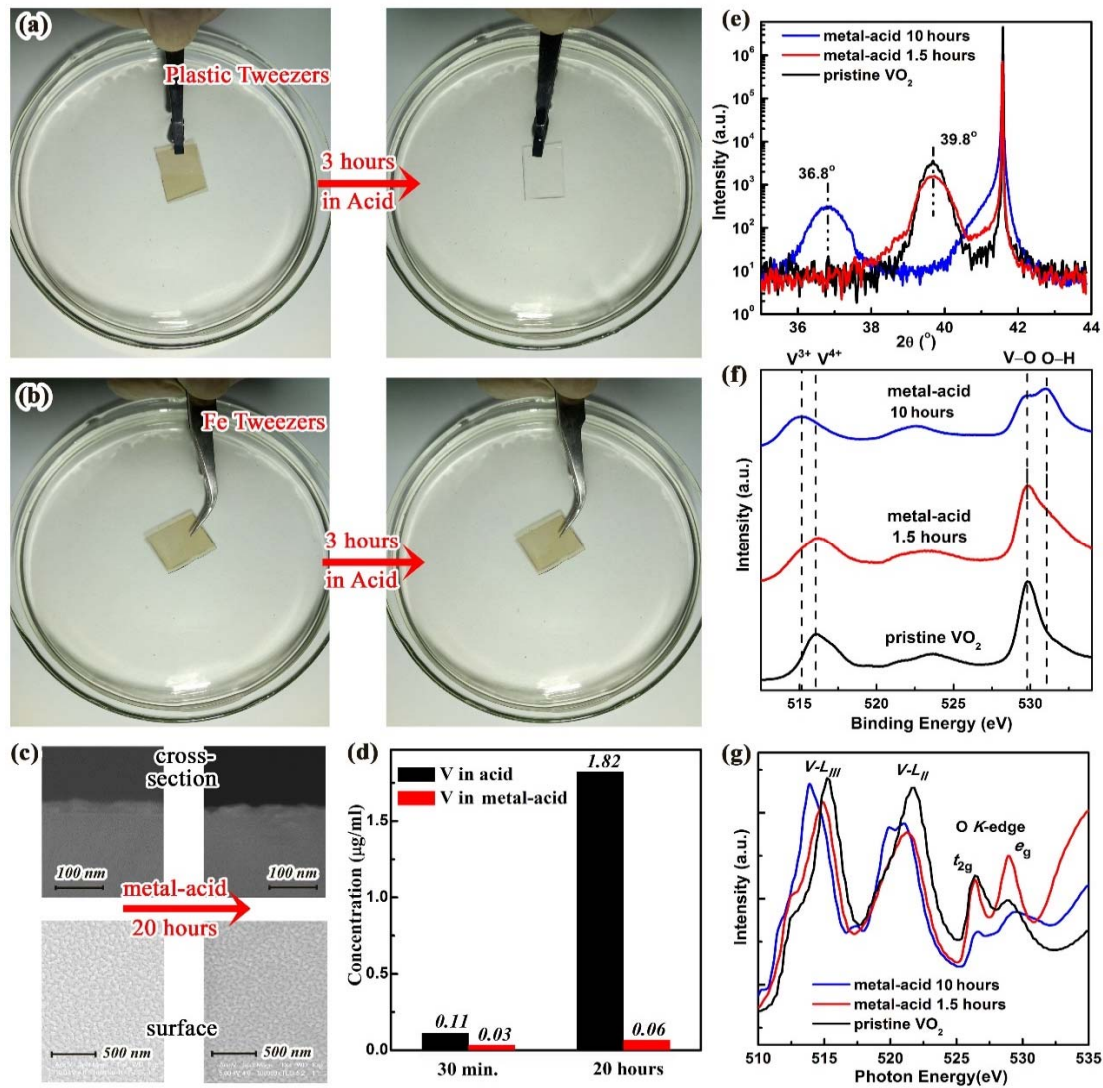

Figure 1 **The metal attached VO$_2$ film show anti-corrosion property in acid solution due to hydrogen intercalation.** (a, b) The VO$_2$ film on Al$_2$O$_3$ substrate held by a plastic tweezers was dissolved by 2% H$_2$SO$_4$ acid in 3 hours, while a steel (Fe) tweezers attachment made the film intact in acid, showing pronounced anti-corrosion ability; (c) The SEM images on the cross-section and surface of the VO$_2$ film being treated by metal-acid for 20 hours, showing that the VO$_2$ film maintained the unchanged thickness and surface morphologies; (d) Trace element analysis monitored V$^{4+}$ concentrations in solution changing from 0.11μg/ml to 1.82 μg/ml after 30 minutes to 20 hours acid treatment, suggesting the VO$_2$ film corrosion in acid. While the very low V$^{4+}$ concentration at 0.03~0.06 μg/ml for 30 minutes to 20 hours demonstrated no corrupted VO$_2$ crystal in acid if metal was attached. (e, f, g) The XRD, XPS, XANES characterizations for the pristine VO$_2$ sample as well as the samples with metal-acid treatment for 1.5 and 10 hours. The pronounced (020) XRD peak shift from 39.8º to 36.7º, the increased V$^{3+}$ and O-H XPS signals, and enhanced

$e_g/t_{2g}$ XANES signal ratio (reflecting the variation of electron occupancy) along with increasing metal-acid time, all indicated lattice changes and O-H bonds formations due to light and heavy hydrogenations in regards to short and long metal-acid treating time. These agree well with the XRD, XPS, and XANES results of the $VO_2$ samples hydrogenated by conventional noble-metal (Au) catalysis at the temperature of 120 ºC (Supplementary Information Fig. S3).

The anti-corrosion ability of metal-acid treated $VO_2$ should be ascribed to hydrogenation, since conventional hydrogenated $VO_2$ assisted by Au or Pd catalyst is also very stable in acid solution (Supplementary Information Fig. S2). For the case of metal-acid treated $VO_2$ film, protons in acid solution reside into $VO_2$ crystal with the help of attached metal to form O–H bonds instead of destroying the $VO_2$ lattice, which then prohibit the attack of $H^+$ to oxygen atoms. The X-ray diffraction (XRD) spectra in Fig. 1e show the dynamics shifts of (020) diffraction peak from 39.8º to 36.7º after the metal-acid treatment, which agree with XRD curves of lightly and heavily hydrogenated $VO_2$ through conventional noble-metal catalysis at high temperature (Supplementary Information Fig. S3). These suggest slightly expanded cell volumes due to H-incorporation. Fig. 1f presented the XPS measurement results, showing the conversion from $V^{4+}$ to $V^{3+}$ state due to H intercalation. The variations of O$1s$ peak at ~531.6 eV for the O–H species further confirmed H-incorporation in $VO_2$ after metal-acid treatment. Furthermore, the XANES spectra in Fig. 1g show the V L-edge curves shifting continuously to lower energy, indicating the polarized charge in V atoms and the evolution of valence state from $V^{4+}$ to $V^{(4-\delta)+}$ or even to $V^{3+}$ state. The polarized charge effect was also inferred from the O K-edge signal. After metal-acid treatment, the relative intensity ratio of the $t_{2g}$ and $e_g$ peaks decreased substantially, reflecting the variation of electron occupancy due to electron doping. All of these spectroscopic variations induced by metal-acid treatment with 1.5 and 10 hours, agree well with corresponding measurements on lightly and heavily hydrogenated $VO_2$ through conventional catalysis techniques (Fig. S3), respectively. These together with above corrosion-protection tests, demonstrated that such metal-acid treatment indeed created H-doping in the $VO_2$ film.

To further explore the effect of metal attachment, a tiny Cu particle (~1.0mm in diameter) was attached to the center of one 2-inch M-VO$_2$/Al$_2$O$_3$(0001) epitaxial film, which were immersed into 2% H$_2$SO$_4$ solution. It was observed in Fig. 2a that, the bare VO$_2$ film with yellowy color was dissolved within 1.5~3 hours, indicating the complete corrosion of VO$_2$ layer. In sharp contrast, although the ~1.0mm copper particle was contacted to the VO$_2$ film with a very small area, it protected the whole 2-inch wafer from acid corrosion. In addition, even if we took away the Cu particle after the treatment, the film was still stable in acid solution (Supplementary Information Fig. S4).

It is known that hydrogenation for VO$_2$ can induce MIT at room temperature[2, 8], *i. e.* hydrogenation converted the insulated M-VO$_2$ to be metallic (Supplementary Information Fig. S5). Starting from the original insulating M-VO$_2$ film (Fig. 2b), the above Cu-acid treatment lowered down its surface resistance for ~3 orders of magnitude (Fig. 2c). Applying heat the sample in air at 120°C to remove the intercalated hydrogens within half an hour, the film was recovered back to the insulated phase (Fig. 2d), which is consistent with the results of hydrogenated samples through conventional catalysis (Supplementary Information Fig. S6). From the resistance distribution map in Fig. 2b, one should note again that the ~1.0mm copper particle accomplished MIT for the whole 2-inch VO$_2$ wafer, including the center copper-covered area where copper was gradually eliminated by acid (Fig. 2a). These are of great advantages for achieving complete and clean H-doping materials, as the conventional catalysis-based technique is suffered to the limited hydrogenation area covered by catalysts (Fig. S5b), as well as the difficulty to remove metal catalysts after hydrogenation.

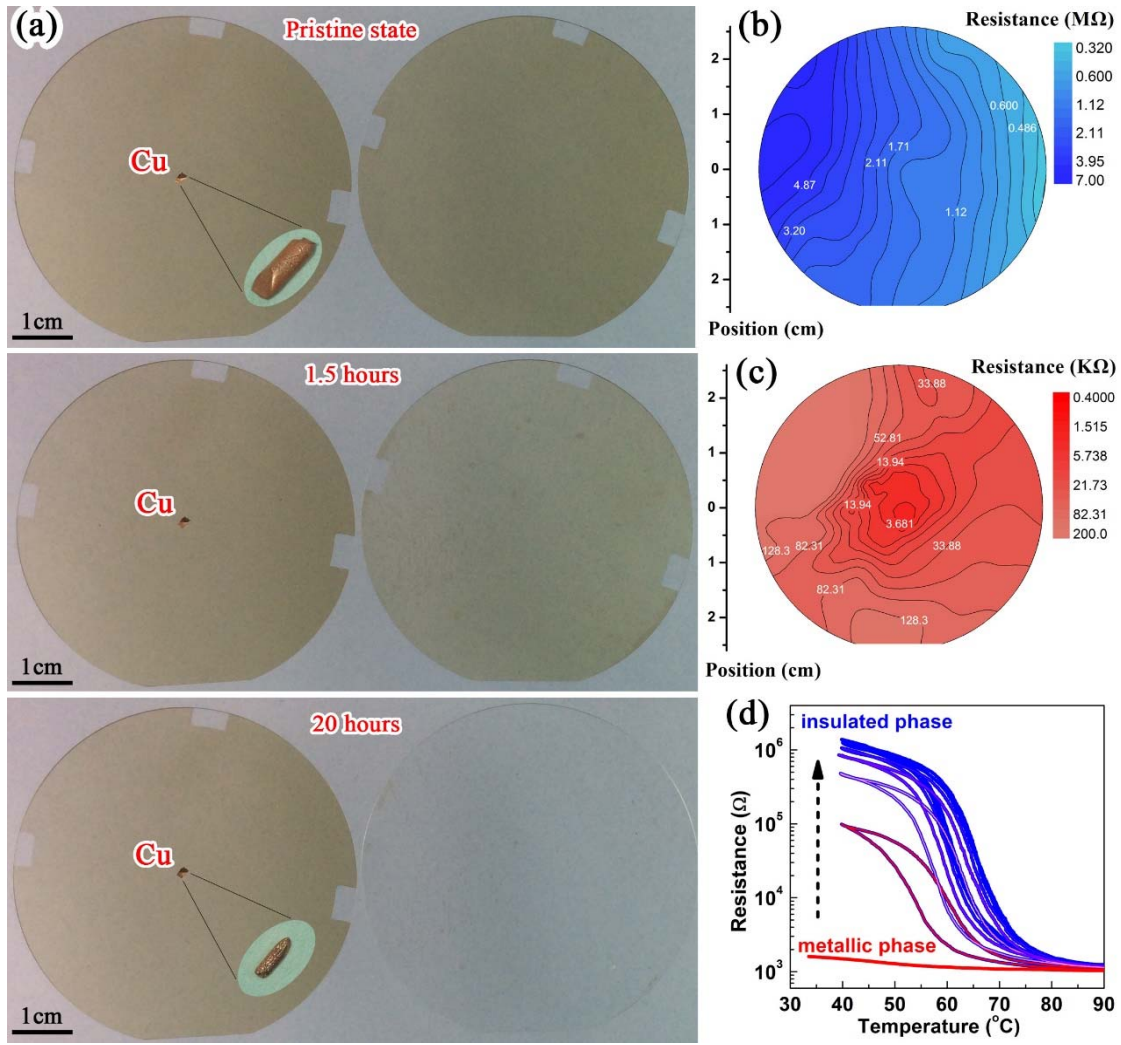

Figure 2 **A tiny Cu particle protects a 2-inch VO$_2$ crystal film from corrosion in acid solution and induces the phase transition due to hydrogenation**. (a) The corrosion of two 2-inch VO$_2$/Al$_2$O$_3$ wafers immersed in 2% H$_2$SO$_4$ acid solution. The sample with a tine copper (Cu) particle (~1.0mm) attached on the center surface exhibited pronounced anti-corrosion ability, while the bare VO$_2$/Al$_2$O$_3$ film is completely corroded within 1.5~3 hours, leaving the transparent Al$_2$O$_3$ substrate. (b) The resistivity mapping for the 2-inch pristine VO$_2$ film; (c) The resistivity mapping for the metal(Cu)-acid treated VO$_2$. For the whole 2-inch wafer, the surface resistance was dropped for ~3 orders of magnitude in comparing to the pristine film, reflecting the MIT to M-VO$_2$ by hydrogenation. (d) The R-T measurement in air for the metal(Cu)-acid treated M-VO$_2$ with heat to remove the doped hydrogens. Along with the pronounced hysteresis R-T curve, the metallic sample gradually recovers to the initial insulated M-phase VO$_2$.

We thus move forward to examine the underlying mechanism of the metal-acid induced VO$_2$ hydrogenation. In Fig. 3a, it is found that active metals including Al, Cu, Ag, Zn or Fe (the pictures of Zn and Fe were not shown here) could all induce hydrogenation and thereby protect VO$_2$ from corrosion in acid. In sharp contrast, the noble and relatively stable metal of Au and Pt could not. Here metals and VO$_2$ actually formed the typical metal-semiconductor interfaces, forming the so-called Schottky Contacts. Theoretical investigations at the first-principle level were performed to compute the workfunctions of metals and M-VO$_2$ (Supplementary Information Fig. S7), and listed together with reported experimental values[27] in Fig. 3b. Due to the workfunction differences, metals with higher electric Fermi level ($E_F$) would donate electrons to the interfaced semiconductor with lower $E_F$ (Fig. 3c). As expected, Al, Cu, Ag, Zn have lower workfunction than M-VO$_2$ (Fig. 3b), so that one (1×1) VO$_2$ unit could extract 0.47~2.50 e$^-$ from metals (Fig. 3d, Supplementary Information Fig. S8 and Table. S1). On the other hand, Au and Pt metal-attachments with higher workfunction than VO$_2$ induced nearly no extra electrons in the interface M-VO$_2$.

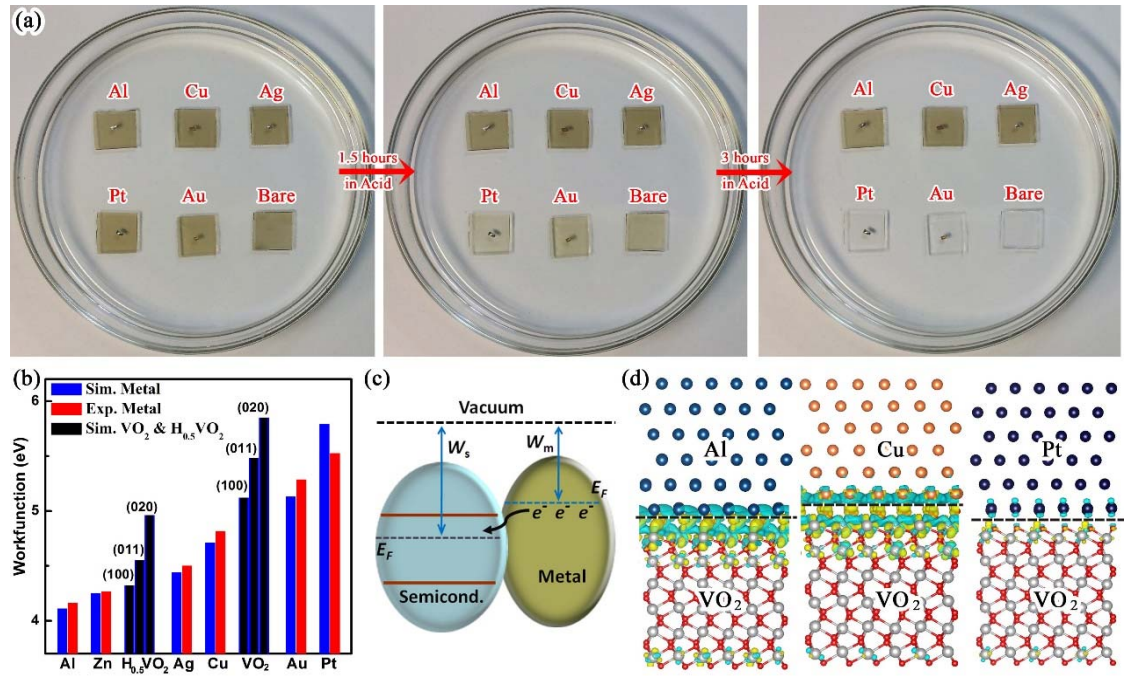

Figure 3 **Different hydrogenation effect induced by different acid-metal treatments to VO$_2$ film due to the workfunction differences.** (a) Active metals including Al, Cu, Ag can protect M-VO$_2$ from corrosion in 2% H$_2$SO$_4$ acid solution, while the M-VO$_2$ were dissolved in 1.5~3 hours if attached to stable metals of Pt and Au. (b) Computed and experimental[27] workfunction ($W_F$) values

for metals, M-VO$_2$, and lightly hydrogenated H$_{0.5}$VO$_2$, with the order of Pt > Au > VO$_2$ > Cu > Ag > H$_{0.5}$VO$_2$ > Zn > Al. XRD measurements in Fig. 1e identified our VO$_2$ sample with the (020) facet. Here we focused on three facets of (020), (011), and (100), among which the latter two are the most stable surfaces for M-VO$_2$. See calculation details in Supplementary Information Fig. S7. (c) Schematic depiction of electrons flowing from metal with a higher Fermi level (i.e., lower work function $W_m$) to semiconductor with a lower Fermi level (i.e., higher work function $W_s$) at the interface. (d) Computed differential charge distribution at Al/Cu/Pt−VO$_2$(020) interfaces, showing that active metals (Al and Cu) donate effective electrons to VO$_2$. Green and yellow bubbles represent hole and electron charges, respectively. Gray, red, brown beads stand for V, O, Cu atoms, respectively.

Consequently, the extra electrons in M-VO$_2$ doped from active metals would drive surrounding acid protons to penetrate. By examining six VO$_2$ surface sites for a proton to adsorb (Inset graph in Fig. 4a and Supplementary Information Fig. S9), we found that more doped electrons led to higher adsorption energies for all sites (Fig. 4a). For instance, on site 1, the proton adsorption energy of 3.68 eV in neutral circumstance was increased to 5.04 eV for a VO$_2$ unit with 4 e$^-$ charge. The diffusion of surface protons into the VO$_2$ crystal could also be promoted by doped electrons (Supplementary Information Fig. S10). Therefore, driving by the electrostatic attraction force, the surrounding protons could penetrate into VO$_2$ to meet electrons, resulting in neutral H intercalation. The incorporation of H in the VO$_2$ crystal then prohibited further attack/adsorption of protons to oxygen, and increased the formation energy required for oxygen vacancy defect (Supplementary Information Fig. S11), leading to anti-corrosion ability in acid solutions. This thus constitutes an electron-proton co-doping strategy, which creates stable neutral H-doping in VO$_2$.

The H-doping then changed the VO$_2$ electronic structures. For a VO$_2$ unit with small H-doping concentration of H$_{0.25}$VO$_2$ (Fig. 4b), the evolutions of electronic structures were reflected by the computed partial density of state (PDOS) of the V-3d orbitals in Fig. 4b. The formation of H-O bonds causes electrons transferring from H to O atoms (Supplementary Information Table S2), which in turn promoted the electron occupancy of V-3d orbitals (Table S2). Such effects were reflected by the up-shifting of Fermi level from the pure $VO_2$ to $H_{0.25}VO_2$ (Fig. 4b). Originally, $VO_2$ exhibited a typical insulating state, with wide energy gap consisting of fully-occupied valence band and empty conduction band. The H-doping then made the conduction band edge states partially occupied, causing the MIT to $H_{0.25}VO_2$.

Furthermore, theoretical simulations explained the contagious hydrogenation process which enabled a ~1.0mm metal particle to convert a 2-inch semiconductor wafer. The work functions of the lightly hydrogenated $H_{0.25}VO_2$ with three facets of (020), (011), (100) are 4.32~4.96 eV, which are all lower than those of pristine $VO_2$ at 5.12~5.85 eV. For any H-doped $VO_2$ parts created by metal-acid treatment, electrons would flow/dope into neighboring unhoped $VO_2$ with lower Fermi level (Fig. 4c and Supplementary Information Fig. S12). The electron-doping soon drove further proton penetrations to the neighboring unhoped $VO_2$.

The contagious spreading of electron-proton co-doping process can thus be described with Fig. 4d: (I) Metal with lower workfunction donates electrons to the interfaced semiconductor due to Fermi level differences, doping extra electrons to the semiconductor; (II) Doped electrons drive surrounding protons in acid solution to penetrate into the semiconductor, creating H-doped structure and causing insulator-to-metal phase transition. (III) The conductive H-doped structures delivery electrons to adjacent un-doped semiconductor parts, triggering the next round of electron-proton co-doping and insulator-to-metal phase transition. (IV) The repeated "electron flowing–proton penetration–phase transition–electron further flowing" cycle is contagious and expanding quickly toward full H-doped material.

It should be noticed that this metal-acid treatment induced hydrogenation in $VO_2$ crystal completely lies on the synergetic electron-proton co-doping route. The tiny Cu particle induced hydrogenation behavior in large-area $VO_2$ film with a ripple effect will be terminated at the liquid level if the $VO_2$ film is partially immersed in acid solution (Supplementary Information Fig. S13).

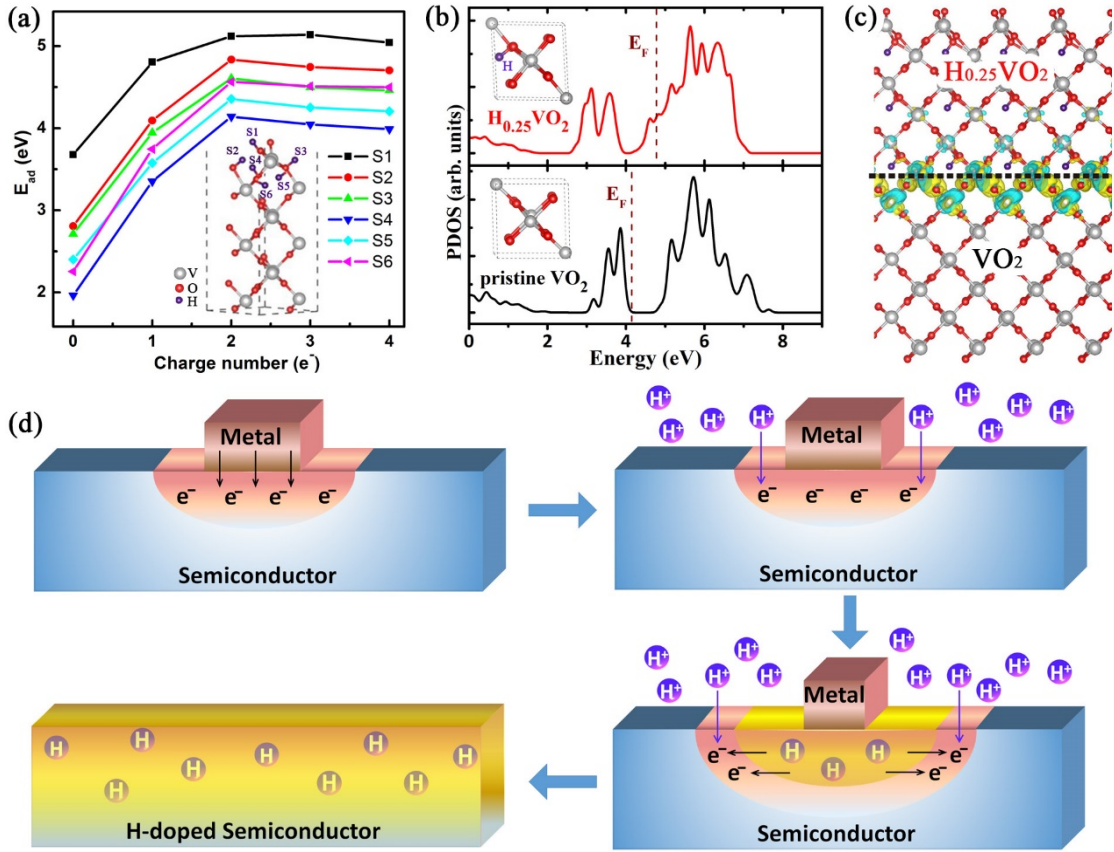

Figure 4 **The mechanism and model for the metal-acid treatment induced hydrogenation in VO$_2$ crystal**. (a) Computed adsorption energies for a proton to six adsorption sites of VO$_2$ (020) surface, increased with the increasing amount of doped electrons. (b) Evolutions of V-3d partial density of state (PDOS), suggest the change of semiconductor band gap in the insulated pristine VO$_2$ to the zero energy gap in H$_{0.25}$VO$_2$. Fermi level is marked with purple dashed lines. (c) Computed differential charge distribution at H$_{0.25}$VO$_2$-VO$_2$ interface, showing each H$_{0.25}$VO$_2$ supercell donated ~2.06 e$^-$ to un-doped VO$_2$. Here green and yellow bubbles represent hole and electron charges, respectively. (d) The schematic illustration of the contagious electron-proton co-doping mechanism with the metal-acid treatment to semiconductor: (1) Electrons flow to semiconductor. (2) Protons penetrate to meet electrons, creating conductive H-doped structure. (3) Electrons flow from conductive H-doped structure to adjacent parts, driving more proton penetration. (4) The repeated "electron flowing–proton penetration–phase transition–electron further flowing" cycle expands toward full H-doping.。

Interestingly, applying more active metals of Al or Zn to hydrogenate VO$_2$ films in acid solution for a long time, the induced metallic state would eventually be transferred into another new insulating state (Supplementary Information Fig. S14). In Fig. 3b, we noticed that Al and Zn actually hold even lower workfunctions than those of lightly hydrogenated $H_{0.25}VO_2$. Therefore, the Al and Zn metal particles would continue to donate much electrons to $H_{0.25}VO_2$, which in turn continued to drive protons penetration until the H-incorporation became saturated. Such transformation behavior of $VO_2$ from the initial insulating phase to the metallic phase and later to insulator, is consistent with very recent findings of the consecutive insulator-metal-insulator transitions induced by increasing H-doping concentration[3]. Amazingly, this metal-acid treatment could be extended to a metal-ions strategy for other elements doping into solids. Replacing the acid solution by Li+ involved polymeric solution, metallic Li-doped $VO_2$ films were obtained (Supplementary Information Fig S15). This demonstrated the universality of the metal-acid induced doping strategy.

In summary, we have proposed a novel electron-proton co-doping strategy to accomplish contagious hydrogenation to $VO_2$ crystal with protons in acid solution. Using non-noble metal (Cu, Ag, Al, Zn, Fe) attachment and diluted acid solution as the electron and proton sources, we achieved the hydrogenation treatment under ambient conditions by injecting both electrons and protons into $VO_2$ film. The resulted H-doping modulated $VO_2$ 3d-orbital electron occupancy and drive consecutive phase transitions between metallic and insulator states. The whole process is extremely efficient as the repeated "electron flowing-proton penetration-phase transition-electron further flowing" cycle is contagious and expanding quickly, so that a tiny metal attachment (~1mm) would convert a very large area of semiconductor (>2 inch). Utilizing the uniformly-distributed protons with controllable concentration in acid, and the well-developed techniques for preparing metal-semiconductor hybrid, this co-doping method would not only help to achieve a peaceable way for hydrogen or Li storage/doping and selective material corrosions, but also stimulate a new way of thinking to develop simple and cost-effective atomic doping technique.

This work was partially supported by the National Basic Research Program of China (2014CB848900), the National Key Research and Development Program of China (2016YFA0401004), the National Natural Science Foundation of China (U1432249, 11574279, 11404095, 21633006), the Fundamental Research Funds for the Central Universities and the research foundation of Key Laboratory of Neutron Physics, China Academy of Engineering Physics (Grant No. 2013BB04). The authors also acknowledge supports from the X-ray diffraction beamline (BL14B1) in Shanghai Synchrotron Radiation Facility, the XMCD beamline (BL12B) and photoelectron spectroscopy beamline (BL10B) in National Synchrotron Radiation Laboratory (NSRL) of Hefei.